\documentclass[conference]{IEEEtran}
\usepackage{algorithm}
\usepackage{algpseudocode}

\usepackage{mathptmx} 
\newcommand{\ignore}[1]{}
\usepackage{fancyhdr}
\usepackage[normalem]{ulem}
\usepackage[hyphens]{url}
\usepackage[pass]{geometry}
\usepackage{multirow}
\usepackage{subcaption}
\usepackage{xspace}
\usepackage{color}
\usepackage{soul}
\usepackage{algorithm}
\usepackage{algpseudocode}

\usepackage{cite}
\usepackage{amsmath,amssymb,amsfonts}
\usepackage{epsfig}

\usepackage{xcolor}

\def\BibTeX{{\rm B\kern-.05em{\sc i\kern-.025em b}\kern-.08em
    T\kern-.1667em\lower.7ex\hbox{E}\kern-.125emX}}

\author{Udit Gupta$^{*\delta}$, Samuel Hsia$^*$, Vikram Saraph$^\delta$, Xiaodong Wang$^\delta$, Brandon Reagen$^\delta$, \\ Gu-Yeon Wei$^*$, Hsien-Hsin S. Lee$^\delta$, David Brooks$^{*\delta}$,  Carole-Jean Wu$^\delta$
\\ \\
$^*$Harvard University\,\,\,\,\,\,\,\,\,\,\,\,\,\,$^\delta$Facebook Inc.\\ \\
ugupta@g.harvard.edu\,\,\,\,\,carolejeanwu@fb.com}

\fancypagestyle{firstpage}{
  \fancyhf{}

  \fancyhead[C]{\normalsize{Under review}} 
  \fancyfoot[C]{\thepage}
}  

\usepackage{booktabs}

\newcommand{\Infra}{DeepRecInfra}
\newcommand{\Sched}{DeepRecSched}

\title{DeepRecSys: A System for Optimizing End-To-End At-scale Neural Recommendation Inference}

\begin{document}
\maketitle
\thispagestyle{firstpage}
\pagestyle{plain}


\begin{abstract}
Neural personalized recommendation is the cornerstone 
of a wide collection of cloud services and products, 
constituting significant compute demand of the cloud 
infrastructure. Thus, improving the execution efficiency 
of neural recommendation directly translates into infrastructure  capacity saving. 
In this paper, we devise a novel end-to-end modeling 
infrastructure, \Infra, that adopts an algorithm and system co-design methodology to custom-design systems for recommendation use cases.
Leveraging the insights from the recommendation
characterization, a new dynamic scheduler, \Sched, is proposed to 
maximize latency-bounded throughput by taking into account characteristics of inference query size and arrival patterns, 
recommendation model architectures, and underlying hardware
systems. By doing so, system throughput is doubled across the eight industry-representative recommendation models. Finally, design, deployment, and evaluation in at-scale production datacenter shows over 30\% latency reduction across a wide variety of recommendation models running on hundreds of machines. 
\end{abstract}




\section{Introduction}

Recommendation algorithms are used pervasively to improve and personalize user experience across a variety of web-services.
%
Search engines use recommendation algorithms to order results, social networks to suggest friends and content, e-commerce websites to suggest purchases, and video streaming services to recommend movies. 
As the sophistication of recommendation tasks increases with larger amounts of better quality data, recommendation algorithms have evolved from simple rule-based or nearest neighbor-based designs~\cite{Sarwar2001CF} to deep learning approaches~\cite{ncf, naumov2019dlrm, cheng2016wide, dinzhou2018deep, dienzhou2019deep, mtwnd}. 

Deep learning-based personalized recommendation algorithms enable a plethora of use cases~\cite{mckinsey}. 
For example, Facebook's recommendation use cases require more than 10$\times$ the datacenter inference capacity 
compared to common computer vision and natural language processing tasks~\cite{kim2018hpca}. 
As a result, over 70\% of machine learning inference cycles at Facebook's datacenter fleets are devoted to recommendation and ranking inference~\cite{gupta2019architectural}. 
Similar capacity demands can be found at Google~\cite{jouppi2017datacenter}, Amazon~\cite{chui2018notes, mckinsey}, and Alibaba~\cite{dinzhou2018deep, dienzhou2019deep}.
And yet, despite their importance and the significant research on optimizing deep learning based AI workloads~\cite{minerva, eie, eyeriss, dla, masr2019} from the systems and architecture community, relatively little attention has been devoted to solutions for recommendation~\cite{kwon2019tensordimm}.
In fact, deep learning-based recommendation inference poses unique challenges that demand unique solutions. 

 \begin{figure}[t!]
    \centering
        \includegraphics[width=\linewidth]{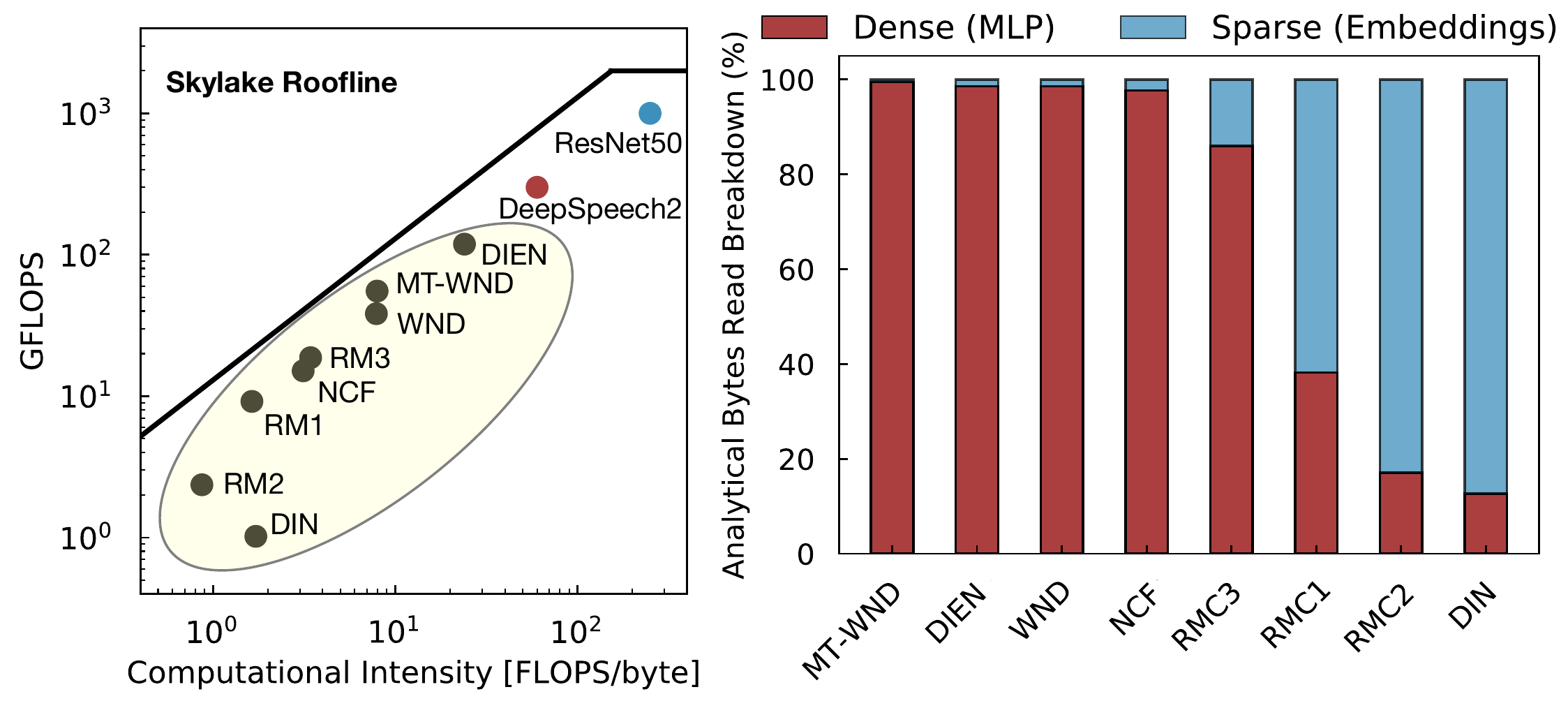}%
    \caption{State-of-the-art recommendation models span diverse performance characteristics compared to CNNs and RNNs. 
    Based on their use case, recommendation models have unique architectures introducing model-level heterogeneity.}
    \label{fig:analytical}%
    \vspace{-1em}
\end{figure}

First, recommendation models exhibit unique compute, memory, and data reuse characteristics.
Figure~\ref{fig:analytical}(a) compares the compute intensity of industry-representative recommendation models\footnote{Section~\ref{sec:infra} describes the eight recommendation models in detail.}~\cite{gupta2019architectural,naumov2019dlrm,dienzhou2019deep,dinzhou2018deep,cheng2016wide,mtwnd,ncf} to state-of-the-art convolutional (CNN) ~\cite{ResNet50} and recurrent (RNN) neural networks~\cite{wu2016google}. 
Compared to CNNs and RNNs, recommendation models, highlighted in the shaded yellow region, tend to be memory intensive as opposed to compute intensive.
Furthermore, recommendation models exhibit higher storage requirements (GBs) and irregular memory accesses~\cite{gupta2019architectural}.
This is because recommendation models operate over not only continuous but also categorical input features.
Compared to the continuous features (i.e., vectors, matrices, images), categorical features are processed by inherently different operations.
This unique characteristic of recommendation models exposes new system design opportunities to enable efficient inference. 

Next, depending on the use case, major components of a recommendation model are sized differently.
This introduces model-level heterogeneity across the recommendation models. 
By focusing on memory access breakdown, Figure~\ref{fig:analytical}(b) shows diversity among recommendation models themselves.
For instance, dense feature processing that incurs {\it regular} memory accesses dominate for Google's WnD~\cite{mtwnd, cheng2016wide}, NCF~\cite{ncf}, Facebook's DLRM-RM3~\cite{gupta2019architectural}, and Alibaba's DIEN~\cite{dienzhou2019deep}. 
In contrast, categorical, sparse feature processing that incurs {\it irregular} memory accesses dominate for other recommendation models such as Facebook's DLRM-RM1/RM2~\cite{gupta2019architectural} and Alibaba's DIN~\cite{dinzhou2018deep}. 
These diverse characteristics of recommendation models expose system optimization design opportunities. 

Finally, recommendation models are deployed across web-services that require solutions to consider effects of executing at-scale in datacenters.
For instance, it is commonly known that requests for web-based services follow Poisson and log-normal distributions for arrival and working set size respectively~\cite{li2016work}.
Similar characteristics are observed for arrival rates of recommendation queries.
However, working set sizes for personalized recommendation queries follow a distinct distribution with heavier tail effects.
This difference in query size distribution leads to varying optimization strategies for recommendation inference at-scale.
Optimizations based on production query size distributions, compared to log-normal, improve system throughput by up to 1.7$\times$ for at-scale recommendation inference.

To enable design optimizations for the diverse collection of industry-relevant recommendation models, 
this paper presents 
{\it \Infra} -- an end-to-end infrastructure that enables researchers to study at-scale effects of query size and arrival patterns.
First, we perform an in-depth characterization of eight state-of-the-art recommendation models that cover commercial video recommendation, e-commerce, and social media~\cite{ncf, gupta2019architectural, dinzhou2018deep, dienzhou2019deep, cheng2016wide, mtwnd}.
Next, we profile recommendation services in a production datacenter to instrument an inference load generator for modeling recommendation queries. 



Built on top of the performance characterization of the recommendation models and dynamic query arrival patterns (rate and size), we propose a hill-climbing based scheduler -- {\it \Sched} -- that splits queries into mini-batches based on the query size and arrival pattern, the recommendation model, and the underlying hardware platform.
\Sched~maximizes system load under a strict tail-latency target by trading off request versus batch-level parallelism. 
Since it is also important to consider the role of hardware accelerators for at-scale AI infrastructure efficiency, \Sched~also evaluates the impact of specialized hardware for neural recommendation by emulating its behavior running on state-of-art GPUs.


The important contributions of this work are:


\begin{enumerate}

  \item This paper describes a new end-to-end infrastructure, \Infra, that enables system design and optimization 
  across a diverse set of recommendation models.
  To take into account realistic datacenter-scale execution 
  behavior, we characterize and integrate query arrival patterns 
  and size distributions observed in production datacenters into 
  \Infra~(Section~\ref{sec:infra}).

  \item We propose a simple, yet effective scheduler -- \Sched --  co-designing the degree of request-
  versus batch-level parallelism based on the dynamic query arrival pattern (rate and size), recommendation model architecture, and service-level latency target (Section~\ref{sec:sched}). 
  Evaluated with \Infra, \Sched~doubles system throughput under strict latency targets.
  In addition,
  we implement and evaluate the design on a production datacenter with 
  live recommendation query traffic, showing significant performance improvement. 
  
  
  \item GPU accelerators can be appealing for recommendation inference, where {\it not all queries are equal}.
  The inflection point varies across the different 
  recommendation models under different system loads and latency targets, showing that 
  \Sched~can dynamically determine the optimal configuration.
  However, compared to CPUs, power efficiency is not optimized in the face of GPUs for recommendation inference (Section~\ref{sec:results}).
  

\end{enumerate}

Systems research for personalized recommendation is still a nascent field. 
To enable follow-on work studying and optimizing recommendation at-scale, we will {\bf open source}\footnote{The open-source implementation will be available upon acceptance of the publication.} the proposed \Infra~infrastructure. 
Open-source \Infra~will include neural personalized recommendation models representative of industry implementations, as well as query arrival rate and size distributions presented in this paper. 

 \begin{figure}[t!]
    \centering
        \includegraphics[width=\linewidth]{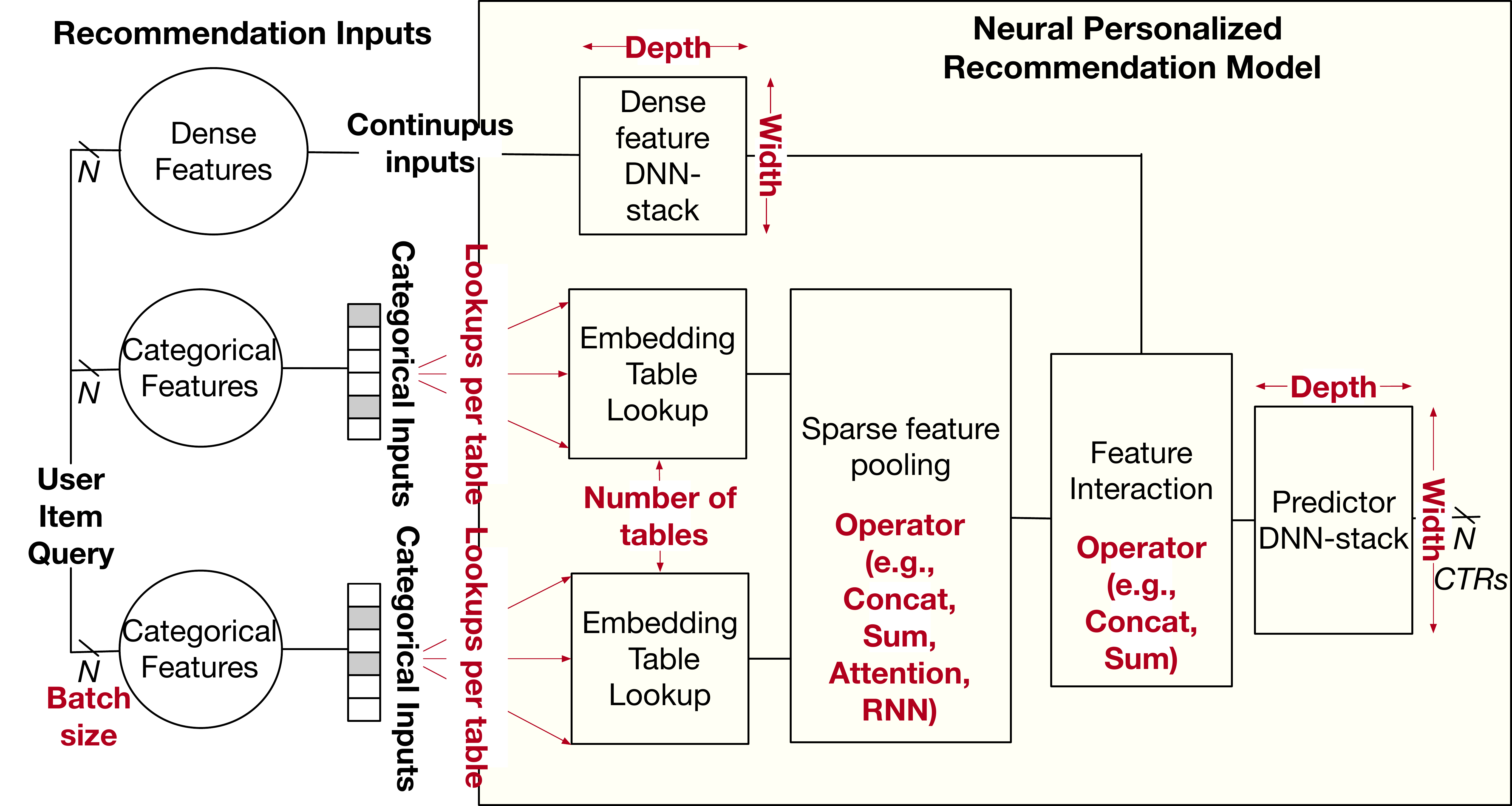}%
    \caption{General architecture of personalized recommendation models.
    Configuring the key parameters (red) yields different implementations of industry-representative models.
    }
    \label{fig:rec_models}%
    \vspace{-1em}
\end{figure}

\section{Neural recommendation models}
\begin{table*}[t!]
\normalsize
\begin{center}
\begin{tabular}{|c|c|c|c|c|c|c|c|}
\hline
       \textbf{Model} & \textbf{Company} & \textbf{Domain} & \textbf{Dense-FC} & \textbf{Predict-FC} & \multicolumn{3}{c|}{\textbf{Embeddings}}  \\ 

        &  &  &  & & Tables & Lookup & Pooling  \\  \hline
       NCF ~\cite{ncf} & - & Movies & - & 256-256-128 & 4 & 1 & Concat \\ \hline 
       Wide\&Deep~\cite{cheng2016wide} &  Google & Play Store & - & 1024-512-256 & Tens & 1 & Concat \\ \hline
       MT-Wide\&Deep~\cite{mtwnd} &  Youtube & Video & - & N x (1024-512-256) & Tens & 1 & Concat \\ \hline

       DLRM-RMC1~\cite{gupta2019architectural} &  Facebook & Social Media & 256-128-32 & 256-64-1 & $\leq$ 10  & $\sim$ 80 & Sum \\ \hline
       DLRM-RMC2~\cite{gupta2019architectural} &  Facebook & Social Media & 256-128-32 & 512-128-1 & $\leq$ 40  & $\sim$ 80 & Sum \\ \hline
       DLRM-RMC3~\cite{gupta2019architectural} &  Facebook & Social Media & 2560-512-32 & 512-128-1 & $\leq$ 10  & $\sim$ 20 & Sum \\ \hline
       DIN~\cite{dinzhou2018deep} &  Alibaba & E-commerce & - & 200-80-2 & Tens  & Hundreds & Attention+FC \\ \hline
       DIEN~\cite{dienzhou2019deep} &  Alibaba & E-commerce & - & 200-80-2 & Tens  & Tens & Attention+RNN \\ \hline
\end{tabular}
\end{center}
\caption{Architectural features of state-of-the-art personalized recommendation models.}
\label{tab:models}
\vspace{-2em}
\end{table*}


Recommendation is the task of personalizing recommending content most relevant to a user based on preferences and prior interactions.
Recommendation is used across many services including search, video and movie content, e-commerce, and advertisements.
%
However, accurately modeling preferences based on previous interactions can be challenging because users only interact with a small subset of all possible items.
For example, for streaming services, a user only watches a small subset of accessible videos.
As a result, unlike inputs to traditional deep neural networks (DNNs), inputs to recommendation models include both \textit{dense} and \textit{sparse} features -- this affects how recommendation models are constructed.

\subsection{Key Components in Neural Recommendation Models}
To accurately model user preference, state-of-the-art recommendation models use deep learning solutions.
Figure~\ref{fig:rec_models} depicts a generalized architecture of DNN-based recommendation models with dense and sparse features as inputs.

{\bf Features.} Dense features describe continuous inputs, such as characteristics of a specific user. 
The dense features are often processed with a stack of MLP layers i.e., fully-connected layers -- similar to classic DNN approaches. 
On the other hand, sparse features represent categorical inputs, such as the collection of products a user has previously purchased or the movies the user has liked
. Since the number of positive interactions for a categorical feature 
is often small compared to the feature's cardinality (all available products), the binary vector representing such interactions ends up very sparse.

{\bf Embedding Tables.} 
Each sparse feature has a corresponding embedding table that is composed of a collection of latent embedding vectors.
The number of vectors, or rows in the table, is determined by the number of categories in the given feature -- this can vary from tens to billions.
The number of elements in each vector, or the column dimension of the table, is determined by the number of latent features for the category representation. This latent dimension is on the order of tens of elements (e.g., 16, 32, or 64). Thus, in total, embedding tables often require storage on the order of tens of GBs.

{\bf Embedding Table Access.} While embedding tables themselves are dense data structures, embedding table operations incur sparse, irregular memory accesses -- especially in the context of personalized recommendation.
Each sparse input is encoded either as one-hot or multi-hot encoded vectors, which are used to index into specific rows of the corresponding embedding table.
The resulting embedding table vectors are combined with a {\it sparse feature pooling} operation such as sum, dot product, or multiplication.
Note that while embedding lookups could be encoded as a sparse matrix-matrix multiplication, it is more computationally efficient to implement the operation as a table lookup followed by a pooling operation. 


{\bf Feature Interaction.} The outputs of the dense and sparse features are combined before being processed by subsequent predictor-DNN stacks. 
Typical operations for feature interaction include concatenation, sum, and averaging. 

{\bf Product Ranking.} The output of the predictor-DNN stacks is the click through rate (CTR) probability for a single user-item pair.
To serve relevant content to users, the CTR of thousands of potential items are evaluated for each user.
All CTR's are then ranked and the top-N choices are presented to the user. 
As a result, deploying recommendation models requires running the models with non-unit batch sizes.



\section{\Infra: At-scale Recommendation} \label{sec:infra}
To better understand the distinct characteristics of and design system solutions for neural recommendation models, we developed an infrastructure, \Infra, to model and evaluate at-scale recommendation inference.
\Infra~is implemented as a highly extensible framework enabling us to consider a variety of recommendation models and use cases.
In particular, \Infra~consists of three key components: (1) a suite of industry-representative recommendation models, (2) industry-representative application level tail latency targets, and (3) real-time query serving based on arrival rates and working set size distributions profiled from recommendation running in a production datacenter.
The following subsections detail these components.


\subsection{Industry-scale recommendation models}

Recent publications from Google, Facebook, and Alibaba present notable differences across their recommendation models~\cite{mtwnd,dinzhou2018deep,dienzhou2019deep,gupta2019architectural,ncf}.
The generalized recommendation model architecture shown in Figure~\ref{fig:rec_models} can be customized by configuring key parameters in order to
realize different implementations of recommendation
services that exhibit a variety of distinct performance characteristics.

\subsubsection{State-of-the-art neural recommendation models}
To capture the diversity, \Infra~composes a collection of eight state-of-the-art recommendation models. 
We describe the unique aspects of the recommendation model architecture below and summarize key parameter configurations for each implementation in Table~\ref{tab:models}.

\begin{itemize}
\item \textbf{Neural Collaborative Filtering (NCF)} is a generalization of matrix factorization (MF) techniques popularized by the Netflix Prize~\cite{koren2009matrix}~\cite{funk2006matrix} with multi-layer perceptrons (MLPs) and  non-linearities. 
Following Figure \ref{fig:rec_models}, NCF only considers one-hot encoded sparse features and does not implement a Dense-FC stack. 
The model comprises four embedding tables --- two for users and two for items --- and a relatively small predictor stack.
Following the embedding table operations, sparse pooling implements a generalized MF (GMF) whose outputs are processed by the final predictor stack of MLPs.





\item \textbf{Wide and Deep (WnD)} considers both \textit{sparse} and \textit{dense} input features.
Deployed in Google's Play Store, WnD uses dense features such as user ages and number of applications installed on a mobile platform. 
Combined, the dense features have dimension of $\sim$1000.
Following Figure~\ref{fig:rec_models}, dense input features in WnD bypass the Dense-FC stack and are directly concatenated with the output of one-hot encoded embedding table lookups.
Finally, a relatively large Predict-FC stack produces an output click-through-rate (see Table~\ref{tab:models}).



 \begin{figure}[t!]
    \centering
        \includegraphics[width=\linewidth]{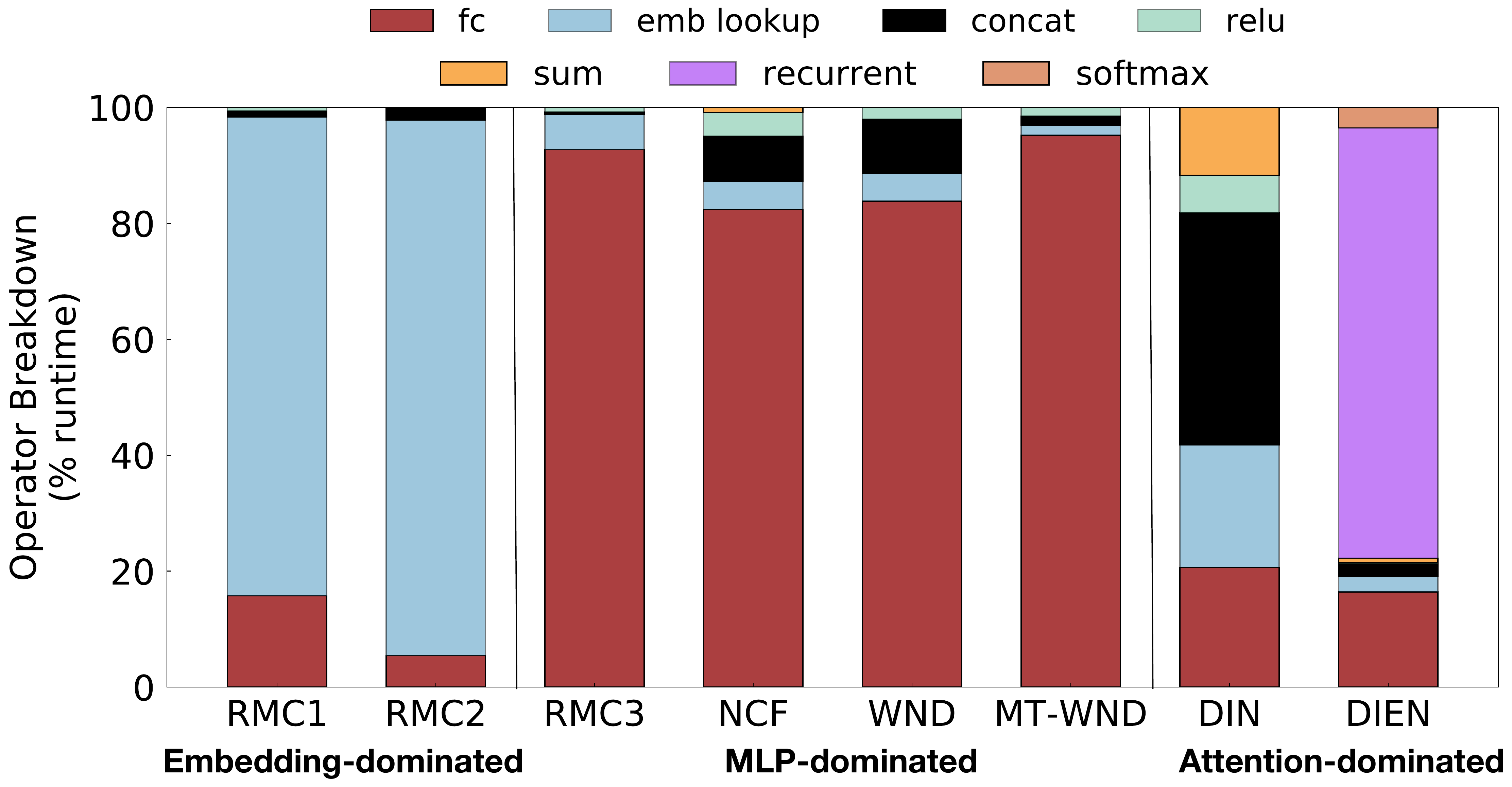}%
    \caption{Operator breakdown of state-of-the-art personalized recommendation models with a batch-size of 64. 
    The large diversity in bottlenecks leads to varying design optimizations.}
    \label{fig:perfCharacterization}%
    \vspace{-1.5em}
\end{figure}

\item \textbf{Multi-Task Wide and Deep (MT-WnD)} extends WnD by evaluating multiple output objectives including predicted click-through rate (CTR), comment rate, likes, and ratings. 
Leveraging multi-objective modeling in personalizing recommendations for users, MT-WnD enables a finer grained and improved user experience~\cite{Baxter1997}.
Building upon WnD, MT-WnD implements $N$ parallel Predict-FC stacks for the different tasks or objectives.




\item \textbf{Deep Learning Recommendation Model (DLRM RMC1, RMC2, RMC3)} is a set of neural recommendation models from Facebook that differs from the aforementioned examples with its large number of embedding lookups~\cite{naumov2019dlrm}. 
In addition, based on Figure~\ref{fig:rec_models}, DLRM first processes the dense features with a DNN-stack.
Based on the configurations shown in ~\cite{gupta2019architectural} varying the number of lookups per embedding table and size of FC layers yield three different architectures, DLRM-RMC1, DLRM-RMC2, and DLRM-RCM3 (see Table~\ref{tab:models}).

\item \textbf{Deep Interest Network (DIN)} uses attention -- implemented as local activation units on top of embedding tables -- to model user interests.
With respect to Figure \ref{fig:rec_models}, DIN does not consider dense input features.
The model comprises tens of embedding tables of varying sizes.
Smaller embedding tables process one-hot encoded user and item features while the larger embedding tables (up to $10^9$ rows) process multi-hot encoded inputs with hundreds of lookups.
The outputs of these multi-hot encoded embedding operations are combined as a weighted sum by a local activation unit (i.e., attention) and then concatenated before being processed by the top predictor stack. 

\begin{figure}[t!]
    \centering
        \includegraphics[width=0.99\linewidth]{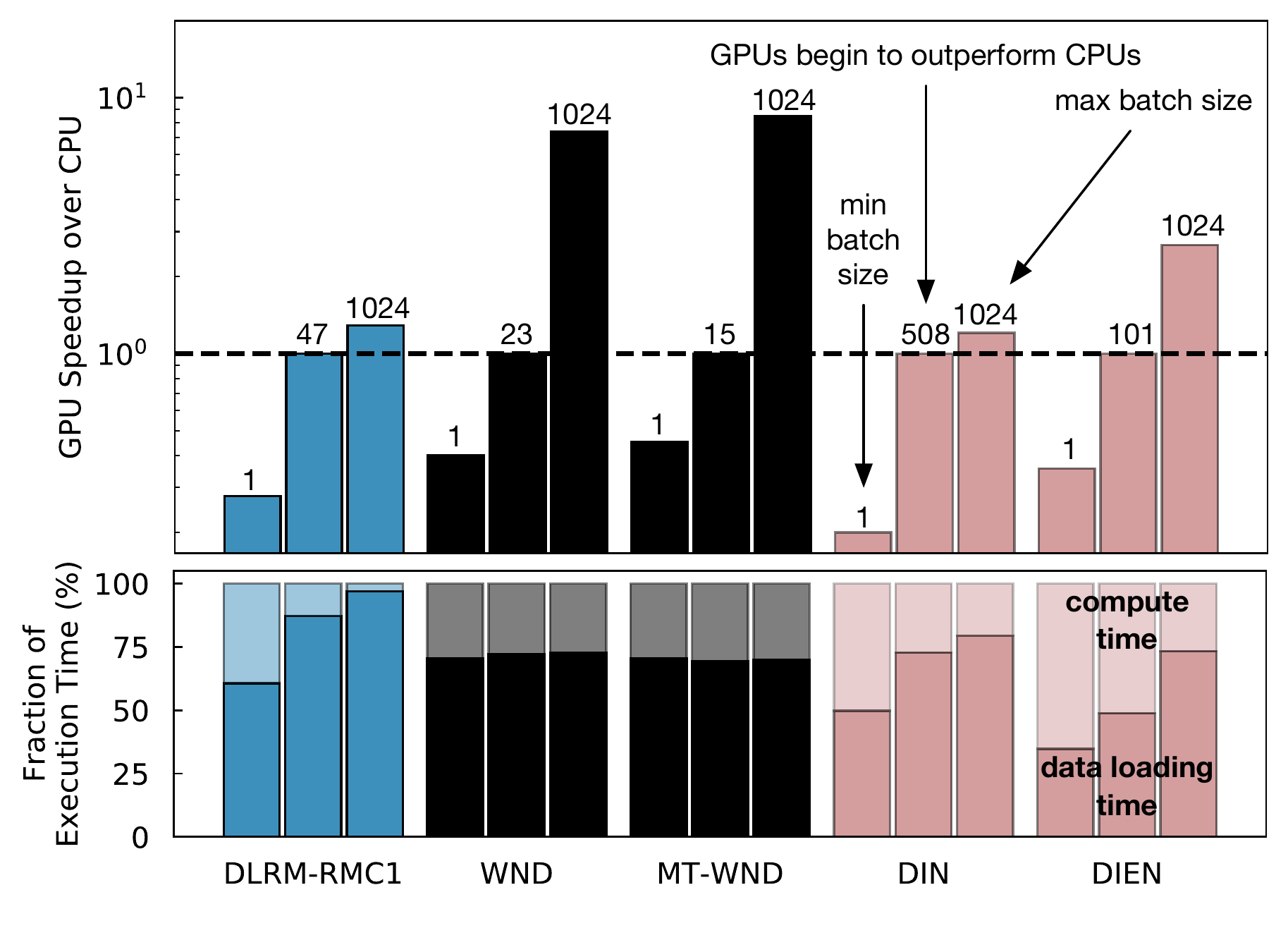}%
    \caption{GPU speedup over CPU for representative recommendation models. 
    GPUs typically have higher performance than CPU at larger batch-sizes (annotated above). 
    The batch-size at which GPUs start to outperform CPUs and their speedup at large batch-sizes varies across models.}
    \label{fig:gpu_speedup}%
\end{figure}



\item \textbf{Deep Interest Evolution Network (DIEN)} captures evolving user interests over time by augmenting DIN with gated recurrent units (GRUs)~\cite{dienzhou2019deep}. 
Inputs to the model are one-hot encoded sparse features.
The output of embedding table operations are processed by attention-based multi-layer GRUs.
The outputs of the GRUs are concatenated with the remaining embedding vectors and processed by a relatively small predictor FC-stack.
\end{itemize}

\begin{table}[t!]
\begin{center}
\begin{tabular}{|c|c|c|c|c|c|c|c|}
\hline
       \textbf{Model} & \textbf{Runtime Bottleneck} & \textbf{SLA target}  \\  \hline
       
       DLRM-RMC1 & Embedding dominated &  100$ms$\\ \hline
       DLRM-RMC2 & Embedding dominated & 400$ms$\\ \hline
       DLRM-RMC3 & MLP dominated & 100$ms$ \\ \hline
       NCF       & MLP dominated & 5$ms$\\ \hline
       WND       & MLP dominated & 25$ms$\\ \hline
       MT-WND    & MLP dominated &  25$ms$\\ \hline
       DIN       & Embedding + Attention dominated & 100$ms$\\ \hline
       DIEN      & Attention-based GRU dominated & 35$ms$\\ \hline

\end{tabular}
\end{center}
\caption{Summarizing performance implications of different personalized recommendation and latency targets used to illustrate design space tradeoffs for \Sched.}
\label{tab:model_perf}
\vspace{-1.5em}
\end{table}



\subsubsection{Operator diversity} \label{sec:op_breakdown}
The apparent diversity of these industry-representative recommendation models leads to a range of performance bottlenecks.
Figure~\ref{fig:perfCharacterization} compares the performance characteristics of recommendation models running on a server class Intel Broadwell, shown as fractions of time spent on Caffe2 operators for a fixed batch size of 64.
As expected, inference runtime for models with high degrees of dense feature processing (i.e., DLRM-RMC3, NCF, WND, MT-WND) is dominated by the MLP layers. 
On the other hand, inference runtime for models dominated by sparse feature processing (i.e., DLRM-RMC1 and DLRM-RMC2) is dominated by embedding table lookups.


Interestingly, inference runtime for attention based recommendation models is dominated by neither FC nor embedding table operations.
For instance, inference run time for DIN is split between concatenation, embedding table, sum, and FC operations. 
This is a result of the attention units, which (1) concatenate user and item embedding vectors, (2) perform a small FC operation, and (3) use the output of the FC operation to weight the original user embedding vector. 
Similarly, the execution time of DIEN is dominated by recurrent layers.
This is a result of fewer embedding table lookups whose outputs are processed by a series of relatively large attention layers. 

\begin{figure}[t!]
    \centering
        \includegraphics[width=0.99\linewidth]{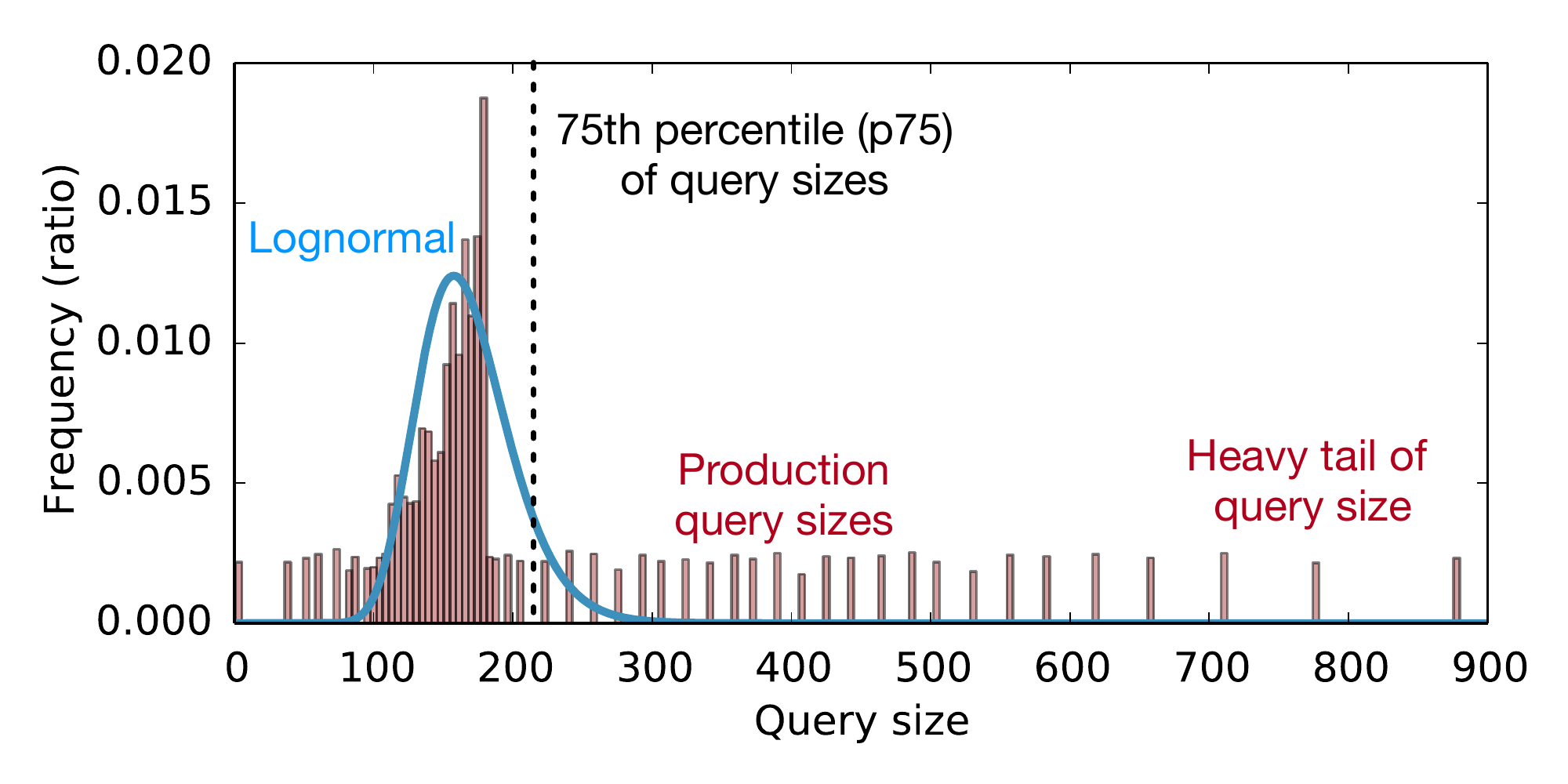}%
    \caption{Queries for personalized recommendation models follow a unique distribution not captured by traditional workload distributions (i.e. normal, log-normal) considered for web-services.
    The heavy tail of query sizes found in production recommendation services leads to unique design optimizations. }
    \label{fig:query_size}%
    \vspace{-1em}
\end{figure}

\begin{figure}[t!]
    \centering
        \includegraphics[width=0.99\linewidth]{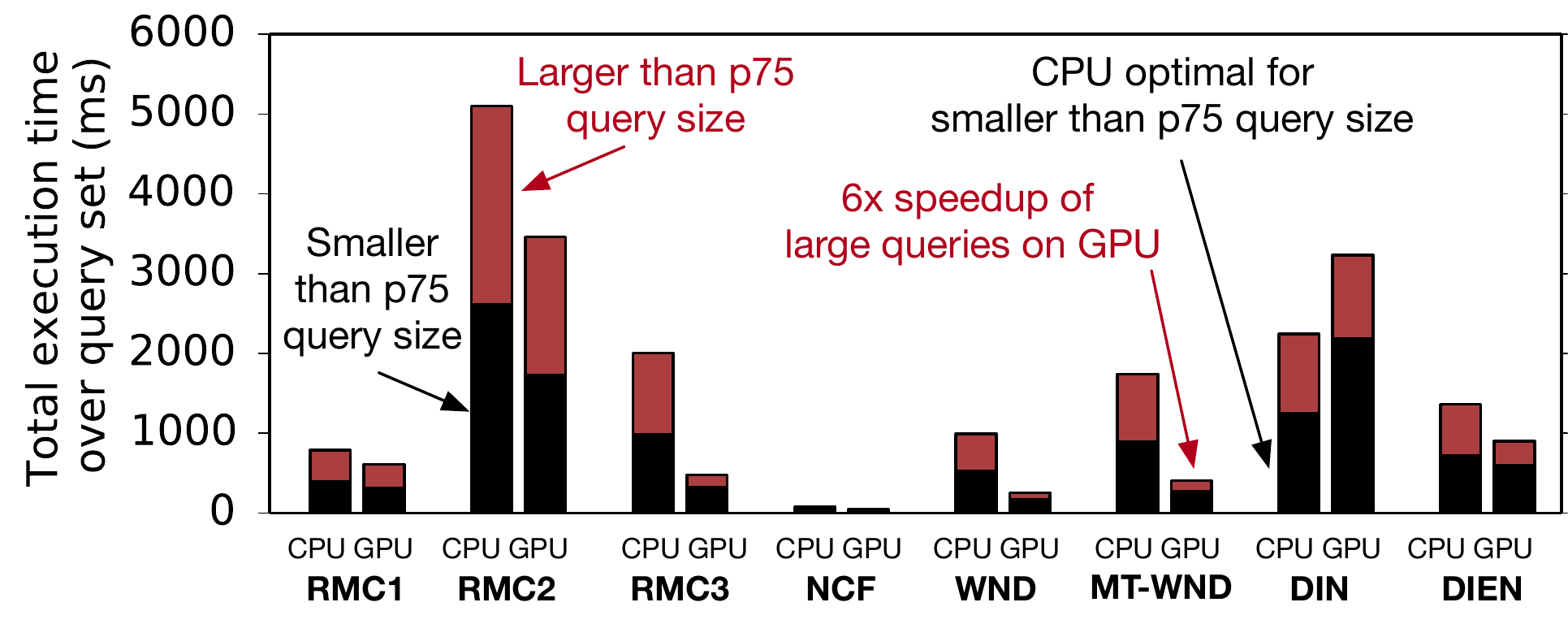}%
    \caption{Aggregated execution time over the query set based on the size distribution for CPU and GPU.
    GPUs readily accelerate larger queries; however, the optimal inflection point and speedup differ across models.}
    \label{fig:query_size_execution}%
    \vspace{-1em}
\end{figure}

\subsubsection{Acceleration opportunity with specialized hardware}
Figure~\ref{fig:gpu_speedup} illustrates the speedup of GPUs over CPUs across different representative recommendation models at various batch sizes.
While GPUs offer higher compute intensity and memory bandwidth, transferring inputs from the CPU to the GPU can consume a significant fraction of time.
For instance, across all batch sizes, data loading time consumes on average 60$\sim$80\% of the end-to-end inference time on the GPU for all recommendation models.
GPUs do, however, provide significant performance benefits at higher batch sizes --- especially for compute intensive models like WnD. Between different classes of recommendation models, (1) speedup at large batch sizes (i.e. 1024) and (2) batch size required to outperform CPU-only hardware platforms vary widely (see Figure~\ref{fig:gpu_speedup}).


\subsection{Service level requirement on tail latency}
Personalized recommendation models are used in many Internet services deployed at a global scale.
They must service a large number of requests across the datacenter while meeting strict latency targets set by the Service Level Agreements (SLAs) of various use cases. 
In this paper, we measure throughput as the number of queries per second (QPS) that can be processed under a $p95$ tail-latency requirement.

\textbf{Diverse set of tail latency targets:} 
We find that the tail latency target varies based on the applications that use recommendation models (e.g., search, entertainment, social-media, e-commerce) and their service-level agreements (SLA).
These differences can result in distinct system design decisions for at-scale recommendation.
Table \ref{tab:model_perf} summarizes the tail latency targets for each of the recommendation models~\cite{gupta2019architectural, mtwnd, cheng2016wide, dinzhou2018deep, dienzhou2019deep}.
For instance, the Google Play store imposes an SLA target of tens of milliseconds on Wide\&Deep~\cite{cheng2016wide, jouppi2017datacenter}. On the other hand, Facebook's social media platform requires DLRM-RMC1, DLRM-RMC2, and DLRM-RMC3 run within an SLA target of hundreds of milliseconds~\cite{gupta2019architectural}.
Alibaba's e-commerce platform requires DIN and DIEN to run within an SLA target of tens of milliseconds (using a collection of CPUs and specialized hardware)~\cite{dinzhou2018deep, dienzhou2019deep}.
In this paper, we use the published targets and profile each model on a server-class Intel Broadwell CPU to set the particular tail-latency target.  Section~\ref{sec:results} then presents system throughput evaluation for a wider range of tail latency targets on optimization strategies and infrastructure efficiency.



\begin{figure}[t!]
    \centering
        \includegraphics[width=\linewidth]{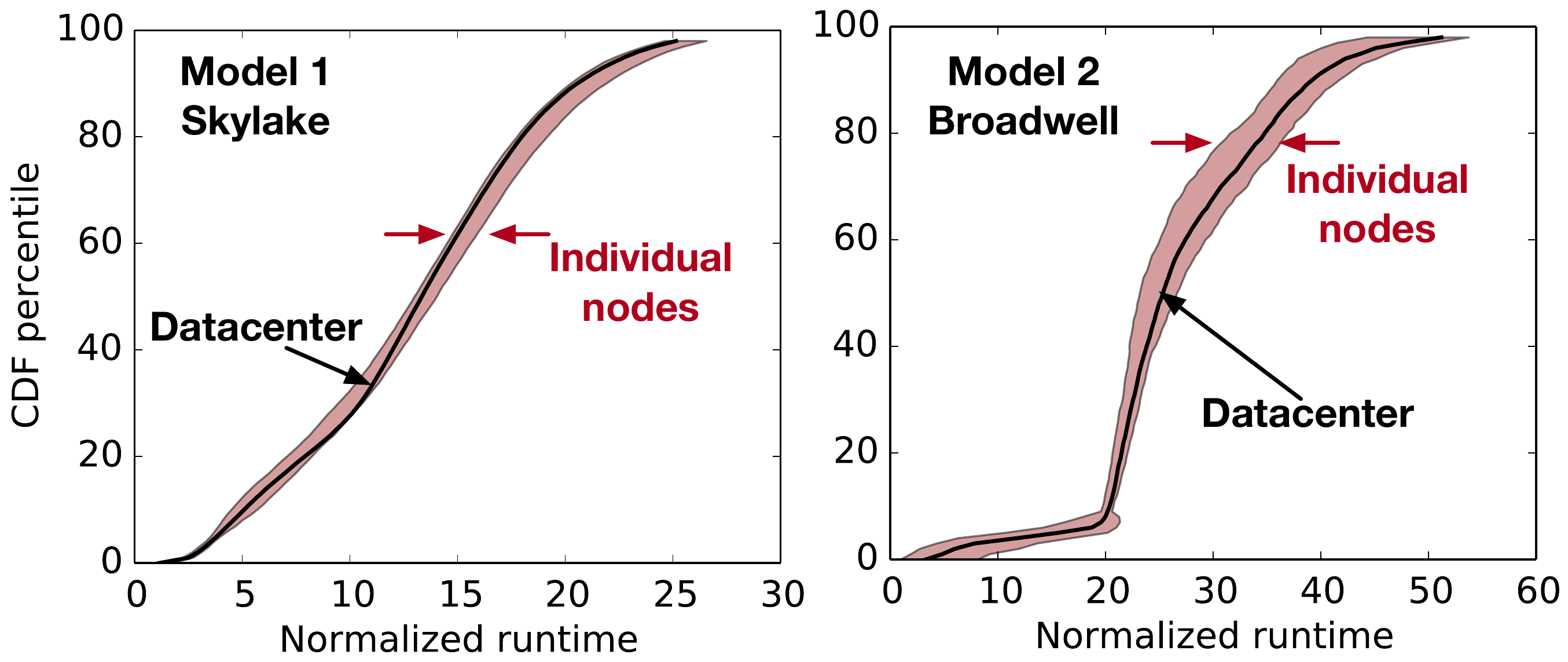}%
    \caption{Performance distribution of recommendation inference at datacenter scale to individual machines.
    Individual machines follow inference distributions, excluding network and geographic effects, at the datacenter scale to within $\sim 9$\%. }
    \label{fig:dc_to_server}%
    \vspace{-1em}
\end{figure}

\begin{figure*}[t!]
    \centering
        \includegraphics[width=0.99\linewidth]{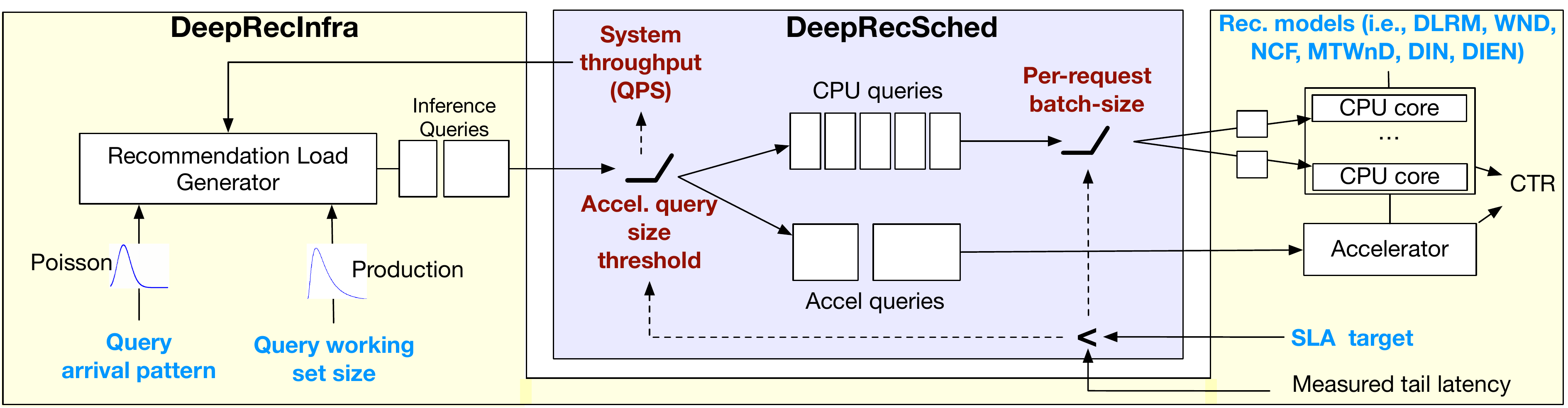}
    \caption{\Infra~implements an extensible framework that considers industry-representative recommendation models, application level tail latency targets, and real-time query serving (rate and size).
    Built upon \Infra, \Sched~optimizes system throughput (QPS) under strict latency targets by optimzing per-request batch-size (request versus batch parallelism) and accelerator query size threshold (parallelizing queries across specialized hardware).
    } \label{fig:deeprecinfra}
    \vspace{-1em}
\end{figure*}


\subsection{Real-Time Query Serving for Recommendation Inference}
It is crucial to model real-time query serving for inference.
\Infra takes into account two important dimensions of real-time query serving: arrival rate and working set sizes.

{\bf Query Arrival Pattern:}
Arrival times for queries for services deployed in the datacenter are determined by the inter-arrival time between consecutive requests. 
This inter-arrival time however can be modeled using a variety of distributions including a fixed value, normal distributions, or Poisson distribution~\cite{hauswald2015sirius,delimitrou2019asplos,mlperf}.
Previous work has shown that, these distributions can lead to different system design optimizations~\cite{li2016work, kandula2009nature }.
Following web-services, by profiling services in a production datacenter, we find that query arrival rates for recommendation services follow a Poisson distribution~\cite{hauswald2015sirius,delimitrou2019asplos,kasture2016tailbench,li2016work,mlperf}.

{\bf Query Working Set Size Pattern:}
Not all recommendation queries are created equally. The size of queries for recommendation inference relates to the number of potential items provided to a user.
Given that the potential number of items to be served depend heavily on the user and their interaction with the web-service, the size of queries varies.
Related work on designing system solutions for web services typically assumes working set sizes of queries follow a fixed, normal, or log-normal distribution~\cite{li2016work}.
However, Figure~\ref{fig:query_size} illustrates that query sizes for recommendation exhibit a heavier tail  compared to canonical log-normal distributions.
As a result, while \Infra's load generator supports a variety of query distributions, the results in the remainder of this paper focus on the query size distribution representative of production datacenter (Figure~\ref{fig:query_size}).

Figure~\ref{fig:query_size_execution} illustrates the execution time breakdown for queries smaller than the p75$^{th}$ size versus larger queries.  
Despite the long tail, the collection of small queries  constitute over half the CPU execution time. {\it 25\% of large queries contribute to nearly 50\% of total execution time.}
This unique query size distribution with a long tail makes GPUs an interesting accelerator target.
Figure~\ref{fig:query_size_execution} shows that, across all models, GPU can effectively accelerate the execution time of large queries, particularly. 
While, offloading the large queries can reduce execution time, the amount of speedup varies based on the model architecture.
The optimal threshold for offloading varies across models, motivating a design that can automatically tune the offloading decision for recommendation inference.



\subsection{Subsampling datacenter fleet with single-node servers}
To serve potentially billions of users across the world, recommendation models are typically run across thousands of machines.
However, it may not always be possible to access and deploy design optimizations across a production-scale datacenter.
We show that we can use a handful of machines to study and optimize tail performance of recommendation inference.
Figure~\ref{fig:dc_to_server} shows the cumulative distribution of two different recommendation models running on server-class Intel Skylake and Broadwell machines.
We find that the datacenter scale performance distribution (black) is tracked by the distribution measured on a handful of machines (red).
The tail-latency trends for recommendation inference across a subset of machines is within 10\% of the performance across machines in a datacenter, representative of larger scale systems.

\subsection{Putting it Altogether}
To study at-scale characteristics of recommendation, it is important to use representative infrastructure.
This includes representative recommendation models, query arrival rates, and query working set size distributions. Putting it all together, we developed \Infra, as depicted in Figure~\ref{fig:deeprecinfra}, by incorporating an extensible load generator to model query arrival rate and size patterns for a diverse set of recommendation models. This enables efficient and representative design and optimization strategies catered to at-scale recommendation.

\section{DeepRecSched design} \label{sec:sched}
In order to consider a variety of recommendation use cases (i.e., model architectures, tail latency targets, real-time query serving, hardware platforms), we design, implement, and evaluate \Sched~on top of \Infra~as shown (Figure~\ref{fig:deeprecinfra}).
By exploiting the aforementioned unique characteristics of recommendation models and real-time query distributions, the proposed \Sched~maximizes system throughput while meeting strict tail latency targets of recommendation services.
Central to \Sched~is the observation that working set sizes for recommendation queries follow a unique distribution with a heavy tail.
Intuitively, large queries, which take the longest to process, limit the throughput (QPS) a system can handle given a strict latency target.
\Sched~addresses this bottleneck with two key design optimizations.
First, large queries are split into multiple requests of smaller batch sizes that are processed by parallel cores.
This requires carefully balancing batch-level and SIMD-level parallelism, cache contention, and the potential increase in queuing delay from a larger number of smaller-sized requests.
Second, large queries are offloaded to specialized AI hardware in order to accelerate at-scale recommendation inference.

\subsection{Optimal batch size varies}
While all queries can be processed by a single core, splitting queries across cores to exploit hardware parallelism, is often advantageous.
Thus, \Sched~splits queries into individual \textit{requests}.
However, this sacrifices parallelism within a request with a decreased \textit{batch size}.

 \begin{figure}[t!]
    \centering
        \includegraphics[width=0.99\linewidth]{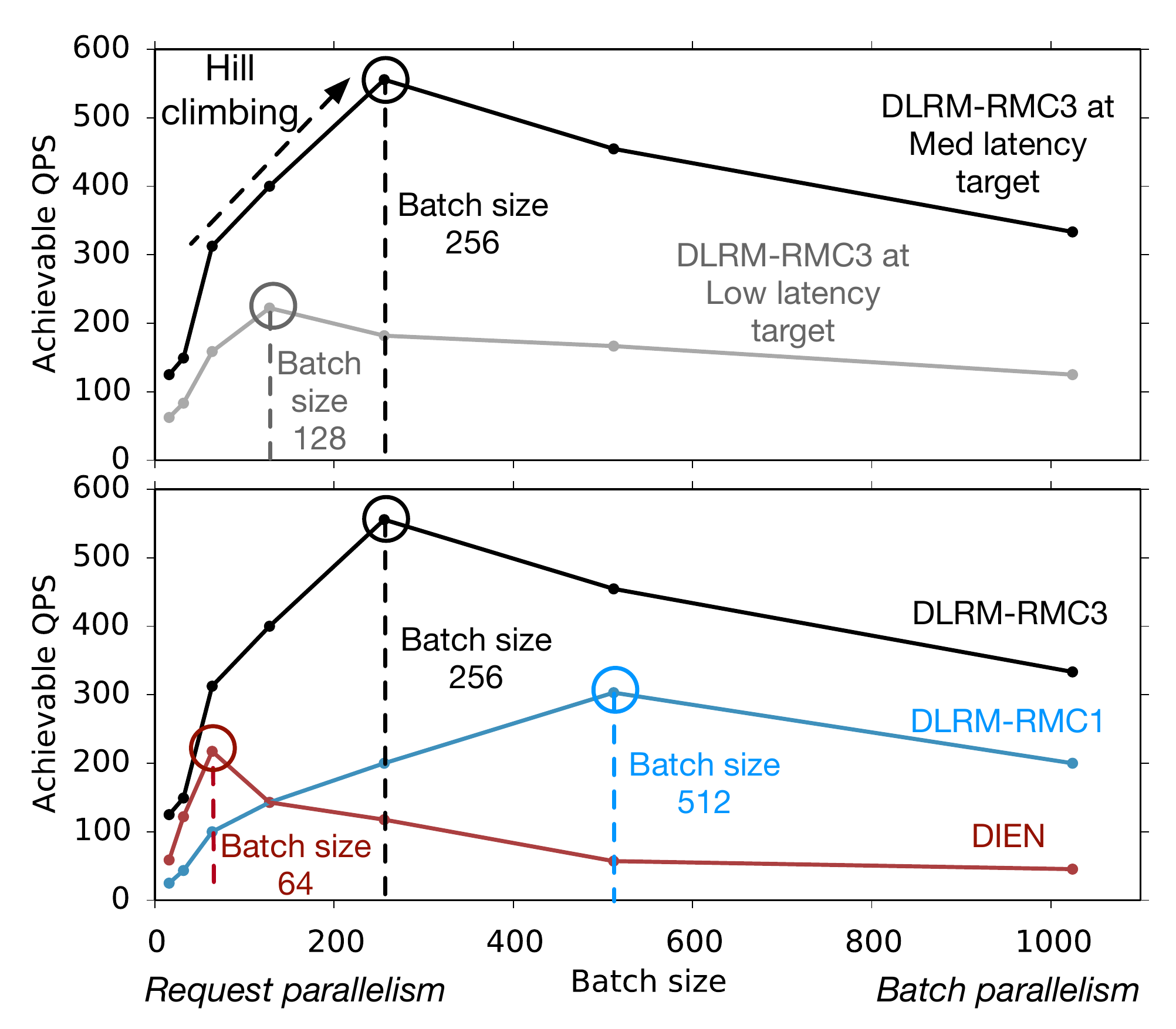}%
    \caption{ Optimal request vs. batch parallelism varies based on the use case.
    (Top) Optimal batch size varies across tail latency targets for DLRM-RMC3.
    (Bottom) Optimal batch size varies across recommendation models i.e., DLRM-RCM2 (embedding-dominated), DLRM-RMC3 (MLP-dominated), DIEN (attention-dominated).}
    \label{fig:cpu_sched_motiv}%
    \vspace{-1em}
\end{figure}

The optimal batch size that maximizes the system QPS throughput varies based on (1) tail latency targets and (2) recommendation models.
Figure~\ref{fig:cpu_sched_motiv}(top) illustrates that, for DLRM-RMC3, the optimal batch size increases from 128 to 256 as the tail latency target is relaxed from 66$ms$ (low) to 100$ms$ (medium).
Furthermore, Figure~\ref{fig:cpu_sched_motiv}(bottom) shows that the optimal batch size for DIEN (attention-based), DLRM-RMC3 (FC heavy), and DLRM-RMC1 (embedding table heavy) is 64, 128, and 256, respectively.


This design space is further expanded considering the heterogeneity of CPUs found in production datacenters~\cite{kim2018hpca}. 
Recent work shows that recommendation models are run across a variety of server class CPUs such as Intel Broadwell and Skylake~\cite{gupta2019architectural}.
Key architectural features across these servers can impact the optimum tradeoff between request- and batch-level parallelism.
First, Intel Broadwell implements SIMD units based on AVX-256 while Skylake implements AVX-512.
Higher batch sizes are typically required to exploit the benefits of the wider SIMD units in Intel Skylake~\cite{gupta2019architectural}. 
Next, Intel Broadwell implements an inclusive L2/L3 cache hierarchy while Skylake implements an exclusive one.
While inclusive cache hierarchies simplify cache coherence protocols, they are more susceptible to cache contention and performance degradation from parallel cores~\cite{jaleel2015high, jaleel2010achieving}.
In the context of recommendation, this can be achieved by trading off request for batch parallelism.
Section~\ref{sec:results} provides a more detailed analysis into the implication of hardware heterogeneity on trading off request- versus batch-level parallelism.



 \begin{figure}[t!]
    \centering
        \includegraphics[width=0.99\linewidth]{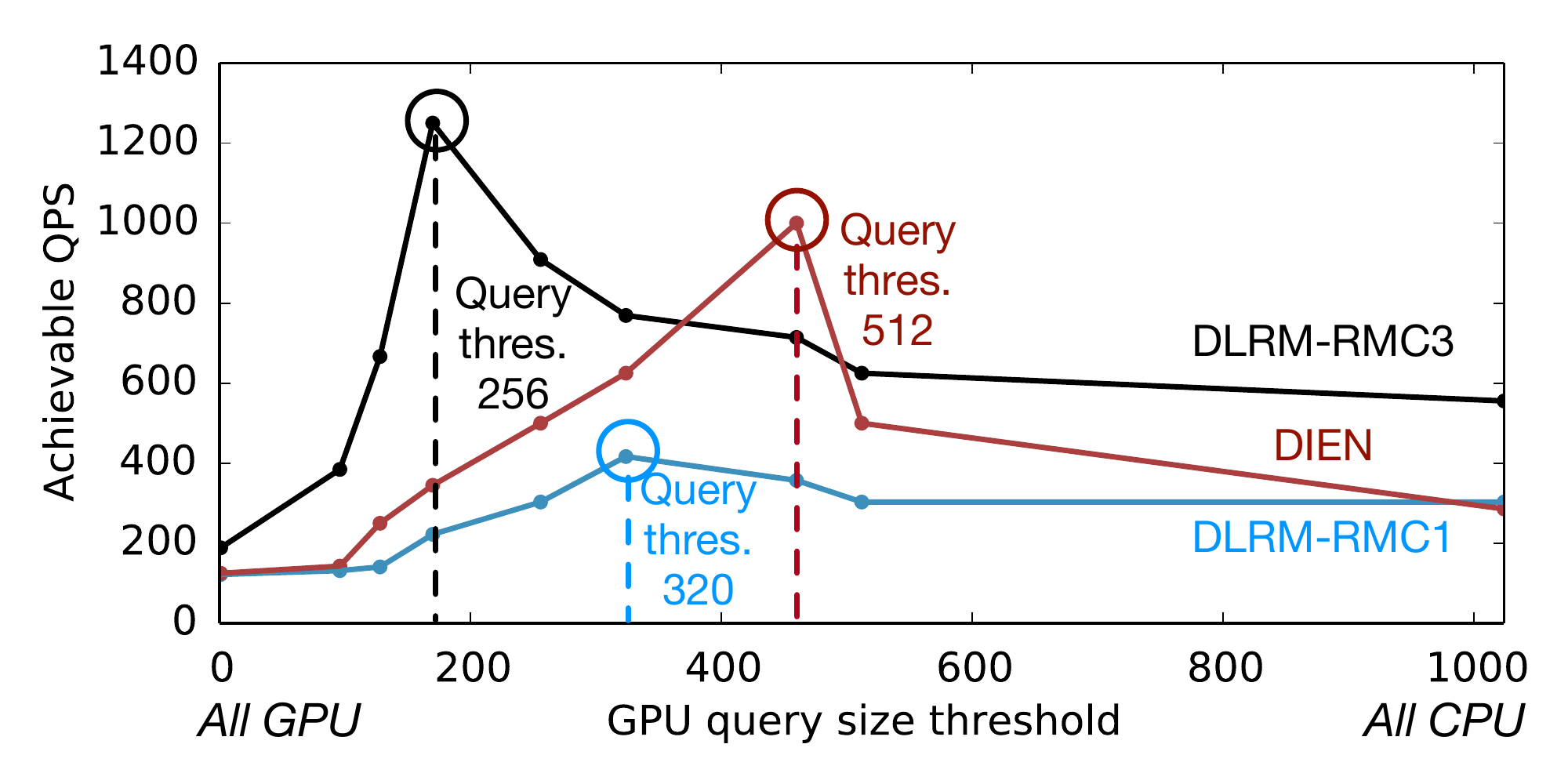}%
    \caption{ The optimal query size threshold, and thus fraction of queries processed by the GPU, varies across recommendation models i.e., DLRM-RMC2 (embedding-dominated), DLRM-RMC3 (MLP-dominated), DIEN (attention-dominated)}
    \label{fig:gpu_sched_motiv}%
    \vspace{-1em}
\end{figure}

 \begin{figure*}[t!]
    \centering
        \includegraphics[width=0.99\linewidth]{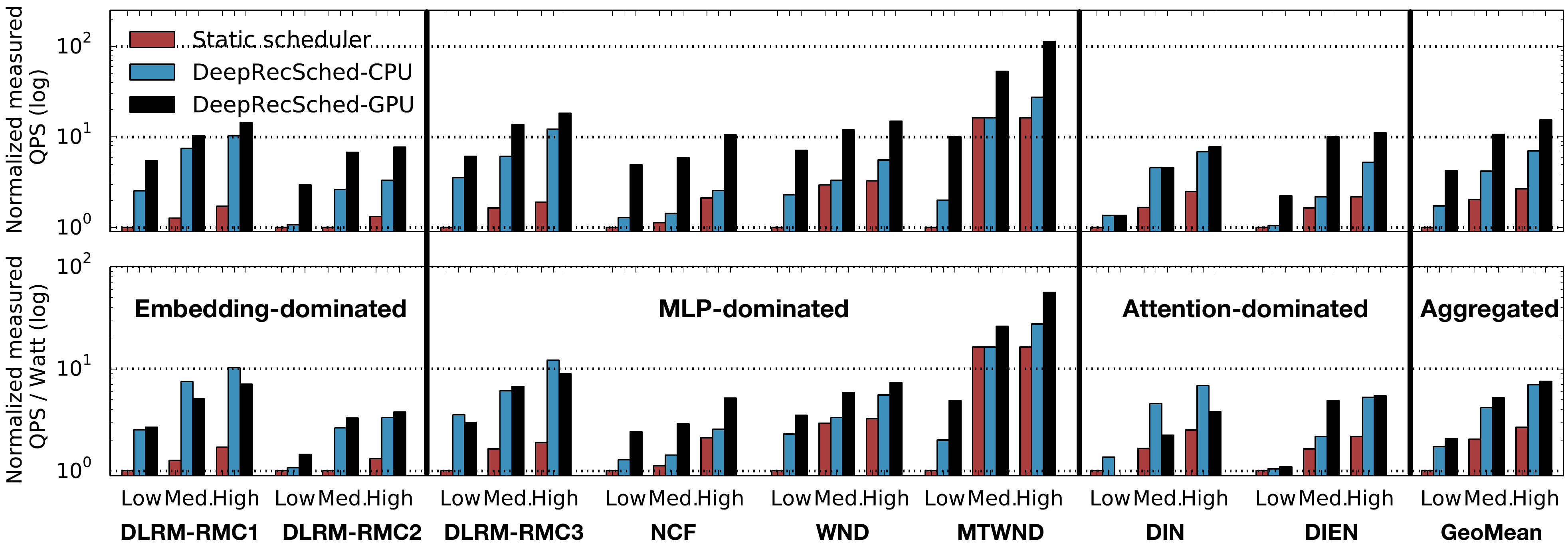}%
    \caption{
    Compared to a static scheduler based on production recommendation services, 
    the top figure shows performance, measured in system throughout (QPS) across a range of latency targets, while the bottom shows power efficiency (QPS/Watt), for \Sched-CPU and \Sched-GPU.
}
    \label{fig:design_opt}%
    \vspace{-1em}
\end{figure*}

\subsection{Leverage parallelism with specialized hardware}
In addition to balancing request- versus batch-level parallelism on general purpose CPUs, in the presence of specialized AI hardware, \Sched~improves system throughput by offloading queries that can best leverage parallelism in the available specialized hardware. 
We evaluate the role of accelerators for at-scale recommendation with state-of-the-art GPUs.
Trading off processing queries on CPUs versus GPUs requires careful optimization.
Intuitively, offloading queries to the GPU incurs significant data transfer overheads. 
To amortize this cost, GPUs often require higher batch sizes to exhibit speedup over general-purpose CPUs, as shown in Figure~\ref{fig:gpu_speedup}~\cite{wei2019benchmarking}. 
Consequently, \Sched~improves system throughput by offloading the largest queries for recommendation inference to the GPU.
This can be accomplished by tuning the \textit{query-size} threshold.
Queries larger than this threshold are offloaded to the GPU while smaller ones are processed by the CPU cores.
Figure~\ref{fig:gpu_sched_motiv} illustrates the impact of query-size threshold (x-axis) on the achievable QPS (y-axis) across a variety of recommendation models.
The optimal threshold varies across the three recommendation models, DLRM-RMC3, DLRM-RMC1, and DIEN.
In fact, we find that the threshold not only varies across model architectures, but also across tail latency targets (see Section~\ref{sec:results} for more details). 

\subsection{\Sched}

One option to identify the optimal batch size that balances the effects of batch- and request-level parallelism is to apply a control-theoretic approach. Based on the detailed characterization results observed in Figures~\ref{fig:cpu_sched_motiv} and~\ref{fig:gpu_sched_motiv}, we find that a simple hill-climbing based algorithm can sufficiently find the optimal batch and query request sizes across the variety of recommendation models and hardware platforms.

\Sched~starts with a unit batch-size to serve recommendation inference queries in \Infra~and increases the batch size to improve system throughput until the achievable QPS degrades while also maintaining the target tail latency.
\Sched~then tunes the query-size threshold for offloading recommendation inference queries to specialized hardware. 
Starting with a unit query-size threshold (i.e., all queries are processed on the accelerator), \Sched~applies hill-climbing to gradually increase the threshold until the achievable QPS degrades.
As what Section~\ref{sec:results} later shows, by automatically tuning the per-request batch size and GPU query-size threshold, \Sched~optimizes infrastructure efficiency of at-scale recommendation across a variety of different model architectures, tail latency targets, query-size distributions, and the underlying hardware.

\section{Methodology}

We implement and evaluate \Sched~with \Infra~across a variety of different hardware systems and platforms. We then compare the performance and power efficiency results with a production-scale baseline.

\textbf{\Infra.} As discussed in Section \ref{sec:infra}, \Infra~comprises three key components:
\begin{itemize}
\item Model Implementation: 
We implement all the recommendation models (shown in Table \ref{tab:models}) in Caffe2 with Intel MKL as the backend library for CPUs~\cite{mkl} and CUDA/cuDNN 10.1 for GPUs~\cite{cudnn}.
All CPU experiments are conducted with a single Caffe2 worker and Intel MKL thread, unless otherwise specified.

\item SLA Latency Targets:
Table \ref{tab:model_perf} presents the SLA targets for each recommendation models. 
To explore the design tradeoffs over a range of latency targets, we consider three latency targets for each recommendation model --- {\it Low}, {\it Medium}, and {\it High} --- where Low and High tail latency targets are set to be 50\% lower and 50\% higher than that of Medium, respectively.

\item Real-Time Query Patterns: 
Following Section~\ref{sec:infra}, query patterns in \Infra~are configurable on two axes: arrival rate and  query size. The arrival pattern is fitted on a Poisson distribution whereas the query sizes are drawn from the production distribution (Figure~\ref{fig:query_size}).

\end{itemize}


\textbf{Experimental System Setup.}
To consider the implications of hardware heterogeneity found in datacenter~\cite{kim2018hpca,gupta2019architectural,knightshift}, 
we evaluate \Sched~with two generations of dual-socket server-class Intel CPUs: Broadwell and Skylake.
Broadwell comprises 28 cores running at 2.4GHz with AVX-2 SIMD units and implements an inclusive L2/L3 cache hierarchy. Its TDP is of 120W.
Skylake comprises of 40 cores running at 2.0GHz with AVX-512 SIMD units and implements an exclusive L2/L3 cache hierarchy. Its TDP is of 125 Watts.

To consider the implications of AI hardware accelerators, we extend the design space to take into account a GPU accelerator model based on real empirical characterization. The accelerator performance model is constructed with the performance profiles of each recommendation model across the range of query sizes over a real-hardware GPU --- server-class NVIDIA GTX 1080Ti with 3584 CUDA cores, 11GB of DDR5 memory, and optimized cuDNN backend library (see Figure~\ref{fig:gpu_speedup}).
This includes both data loading and model computation, capturing end-to-end recommendation inference. 





\textbf{Production-scale baseline.}
We compare \Sched~to the baseline that implements a fixed batch size configuration. 
This fixed batch size configuration is typically set by splitting the largest query {\it evenly} across all available cores on the underlying hardware platform.
Given the maximum query size of 1000 (Figure~\ref{fig:query_size}), the static batch size configuration is determined as 25 for a server-class 40-core Intel Skylake.


 \begin{figure*}[t!]
    \centering
        \includegraphics[width=\linewidth]{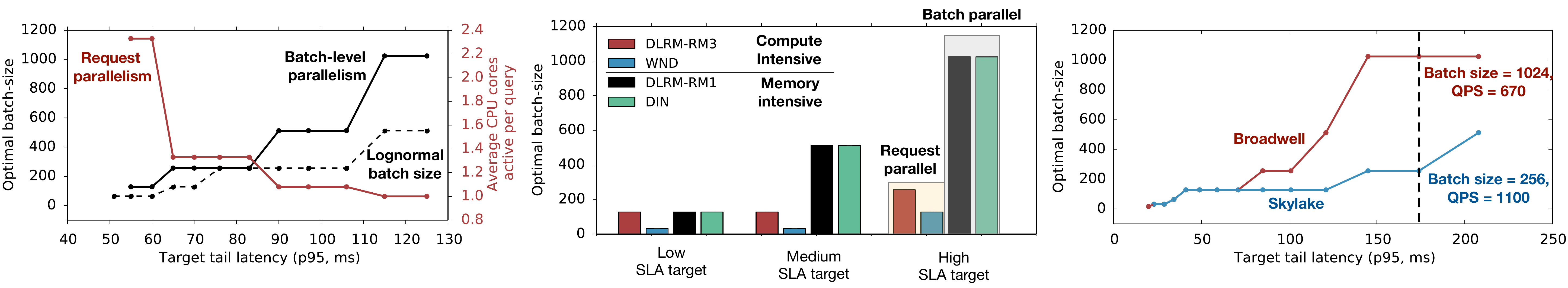}%
    \caption{ 
    Exploiting the unique characteristics of at-scale recommendation yields efficiency improvements given the optimal batch size varies across SLA targets and query size distributions (left), models (middle), and hardware platforms (right). 
    }
    \label{fig:batch_size}%
    \vspace{-1em}
\end{figure*}

\section{\Sched~Evaluation}~\label{sec:results}
This section first presents the overall efficiency improvements of \Sched~over the baseline across all eight state-of-the-art recommendation models using \Infra.
Next, we describe the design tradeoffs and benefits of \Sched~by diving into (1) the tradeoffs in request- versus batch-level parallelism, (2) a case study of demonstrating the benefits of the design optimizations in a real production datacenter, and (3) leveraging parallelization opportunities by offloading requests to specialized hardware. 

\textbf{Performance.}
Figure~\ref{fig:design_opt}(top) compares the throughput performance of \Sched-CPU and \Sched-GPU versus a baseline static scheduler across the three tail latency configurations, all normalized to the measured QPS at the \textit{low} tail latency case of the baseline.
Overall, \Sched-CPU achieves 1.7$\times$, 2.1$\times$, and 2.7$\times$ higher QPS across all models for the \textit{low}, \textit{medium}, and \textit{high} tail latency targets, respectively. 
\Sched-CPU is able to increase the overall system throughput by operating at the optimal batch size configuration. 
Furthermore, \Sched-GPU increases performance improvement to 4.0$\times$, 5.1$\times$, and 5.8$\times$ at the \textit{low}, \textit{medium}, and \textit{high} tail latency targets, respectively.
Thus, parallelizing requests across general-purpose CPUs and specialized hardware provides additional performance improvement for recommendation inference at-scale.

\textbf{Power efficiency.}
Figure~\ref{fig:design_opt}(bottom) compares the QPS-per-watt power efficiency of \Sched-CPU and \Sched-GPU by again normalizing the measured QPS/Watt to the \textit{low} tail latency case of the baseline static scheduler.
Given higher performance under the TDP power budget as the baseline, \Sched-CPU achieves 1.7$\times$, 2.1$\times$, and 2.7$\times$ higher QPS/Watt for all models under the \textit{low}, \textit{medium}, and \textit{high} tail latency targets, respectively.
Aggregated across all models, \Sched-GPU improves the power efficiency improvement  to 2$\times$, 2.6$\times$, and 2.9$\times$ for each latency target.
Compared to the performance improvement, \Sched-GPU provides marginal improvement in power efficiency due to the overhead of GPU acceleration.
In fact, while \Sched-GPU improves system QPS across all recommendation models and latency targets, compared  to  \Sched-CPU, it does not globally improve QPS/Watt.
In particular, the power efficiency improvement of \Sched-GPU is more pronounced for compute intensive models (i.e., WND, MT-WND, NCF). 
For the case of memory intensive models (i.e., DLRM-RMC1, DIN), the power overhead for offloading recommendation inference to GPUs outweighs the performance gain, degrading the overall power efficiency. 
Thus, judicious optimization of offloading queries across CPUs and specialized AI hardware can improve infrastructure efficiency for recommendation at-scale. 

\subsection{Balance of Request and Batch Parallelism}
Compared to the fixed static baseline, \Sched-CPU improves QPS by balancing the request- versus batch-level parallelism across varying tail latency targets, query size distributions, recommendation models, and hardware platforms. 


\textbf{Optimizing across SLA targets.}
Figure~\ref{fig:batch_size}(a) illustrates the tradeoff between request- and batch-level parallelism across varying tail latency targets for DLRM-RMC1.
Under lower, stricter tail latency targets, QPS is optimized at lower batch sizes --- favoring request level parallelism.
On the other hand, at more relaxed tail latency targets, \Sched-CPU finds the optimal configuration to be at a higher batch size --- favoring batch-level parallelism. 
As previously shown in Figure~\ref{fig:design_opt}(top), optimizing the per-request batch size yields \Sched-CPU's QPS improvements over the static baseline across tail latency targets. 


\textbf{Optimizing across query size distributions}
Figure~\ref{fig:batch_size}(a) also shows the optimal batch size, for DLRM-RMC1, varies across query working set size distributions (lognormal and the production distribution).
The optimal batch-size across all tail latency targets is strictly lower for lognormal than the query size distribution found in production recommendation use cases.
This is a result of, as shown in Figure~\ref{fig:query_size}, query sizes in production recommendation use cases following a distribution with a heavier tail.
In fact, applying optimal batch-size configuration based on the lognormal query size distribution to the production distribution degrades the performance of \Sched-CPU by 1.2$\times$, 1.4$\times$, and 1.7$\times$ at low, medium, and high tail-latencies for DLRM-RMC1.
Thus, built ontop of \Infra, \Sched-CPU carefully optimizes request verus batch-level parallelism for recommendation inference in production datacenters.

\textbf{Optimizing across recommendation models.}
Figure~\ref{fig:batch_size}(b) illustrates that the optimal batch size varies across recommendation models with distinct compute and memory characteristics. 
For compute intensive models (e.g., DLRM-RMC3, WnD), system throughput is optimized at lower batch sizes compared to memory intensive models (e.g., DLRM-RMC1, DIN).
At the high SLA targets, DLRM-RMC3 and WnD have an optimal batch size of 256 and 128, respectively. 
This is a result of the compute intensive models being accelerated by the data-parallel SIMD units (i.e., AVX-512 in Intel Skylake, AVX-256 in Intel Broadwell). In addition to leveraging the data-parallel per-core SIMD units, recommendation inference can be further accelerated by processing parallel requests across the chip-multiprocessor (CMP) cores. Running the models with smaller batch sizes result in better request-level parallelism and CMP core utilization. 
On the other hand, DLRM-RMC1 and DIN are optimized at a larger batch size of 1024.
This is because the primary performance bottleneck of models with heavy embedding table accesses lies in the DRAM bandwidth utilization. 
In addition to request level parallelism, memory bandwidth utilization can be improved significantly by running recommendation inference at a higher batch size. 
By exploiting characteristics of the models to optimize the per-request batch size, \Sched-CPU achieves higher QPS for a variety of distinct recommendation models.


 \begin{figure}[t!]
    \centering
        \includegraphics[width=0.99\linewidth]{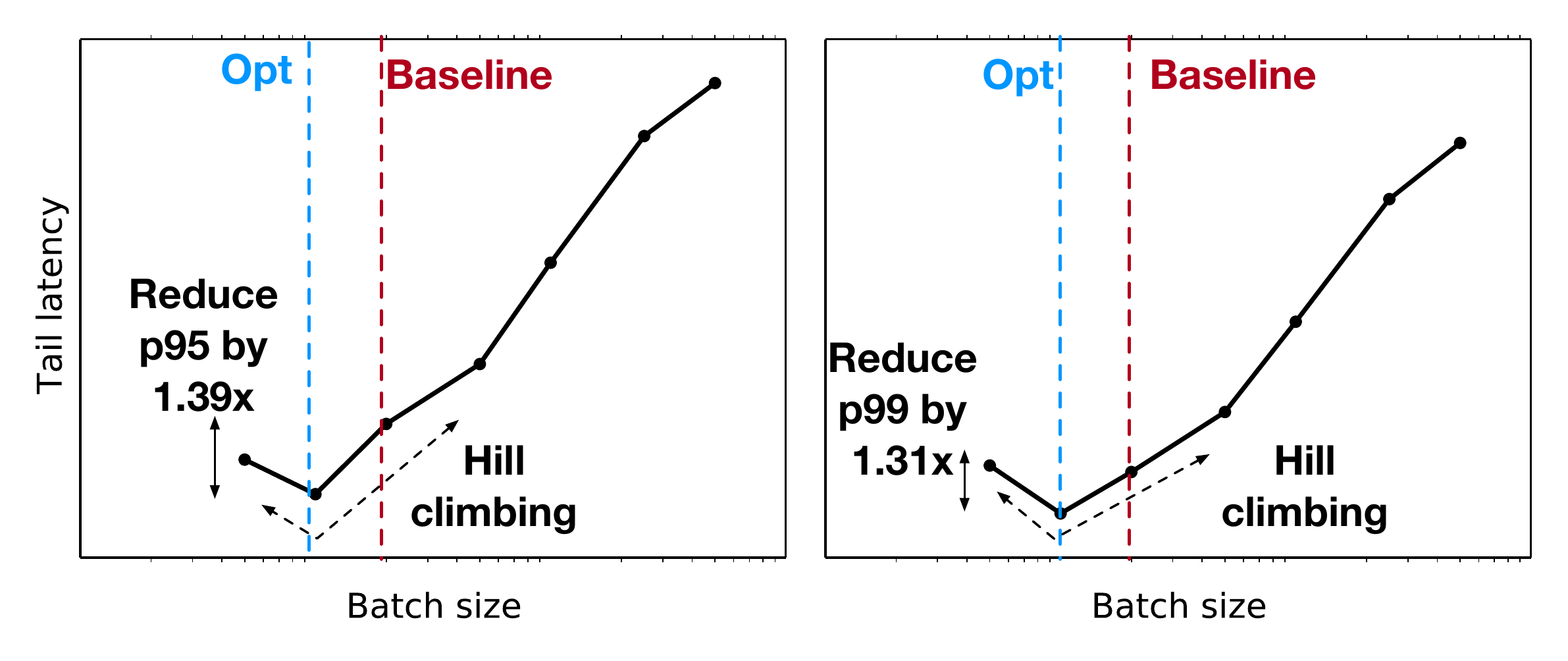}%
    \caption{ 
    Exploiting the request vs. batch-level parallelism optimization demonstrated by \Sched in a real production datacenter improves performance of at-scale recommendation services. 
    Across models and servers, optimizing batch size reduces p95 and p99 latency by 1.39$\times$ (left) and 1.31$\times$ (right). 
    }
    \label{fig:prod_impact}%
    \vspace{-1em}
\end{figure}

\textbf{Optimizing across hardware platforms.}
Figure~\ref{fig:batch_size}(c) shows the optimal batch size, for DLRM-RMC3, varies across server architectures (Intel Broadwell and Skylake machines).
The optimal batch size, across all tail-latency targets, is strictly higher on Intel Broadwell compared to Skylake. For example, at a latency target of 175$ms$, the optimal batch-size on Intel Broadwell and Skylake is 1024 and 256, respectively. 
This is a result of the varying cache hierarchies on the two platforms.
In particular, Intel Broadwell implements an inclusive L2/L3 cache hierarchy while Intel Skylake implements an exclusive L2/L3 cache hierarchy.
As a result, Intel Broadwell suffers from higher cache contention with more active cores leading to performance degradation.
For example, at a latency target of 175$ms$ and per-request batch sizes of 16 (request-parallel) and 1024 (batch-parallel), Intel Broadwell has an L2 cache miss rate of 55\% and 40\% respectively.
To compensate for this performance penalty, \Sched-CPU runs recommendation models with higher batch-sizes --- fewer request and active cores per query --- on Intel Broadwell. 
{\it Overall, \Sched~enables a fine balance between request vs. batch-level parallelism across not only varying tail latency targets, query size distributions, and recommendation models, but also the underlying hardware platforms.}

\subsection{Tail Latency Reduction for At-Scale Production Execution}
In addition to evaluating the proposed design using \Infra, we deploy the proposed design and demonstrate that optimizations translate to higher performance in a real production datacenter.
Figure~\ref{fig:prod_impact} illustrates the impact of varying the batch-size on the measured tail latency of recommendation models running in a production datacenter.
The results are aggregated across a wide collection of recommendation models and server-class Intel CPUs used in the production datacenter fleets.
\textcolor{black}{Experiments are conducted on a cluster consisting of hundreds of machines. These machines are configured to receive a fraction of real-time production traffic.}
To account for the diurnal production traffic as well as intra-day query variability, we deploy and evaluate \Sched over the course of 24 hours.
Compared to the baseline configuration with a fixed batch-size, the optimal batch size provides a 1.39$\times$ and 1.31$\times$ reduction in p95 and p99 tail latencies, respectively.
This reduction in the tail latency can be used to increase system throughput (QPS) serviced by the cluster of machines, as demonstrated by \Sched, translating to datacenter capacity saving.

\begin{figure}[t!]
    \centering
        \includegraphics[width=0.99\linewidth]{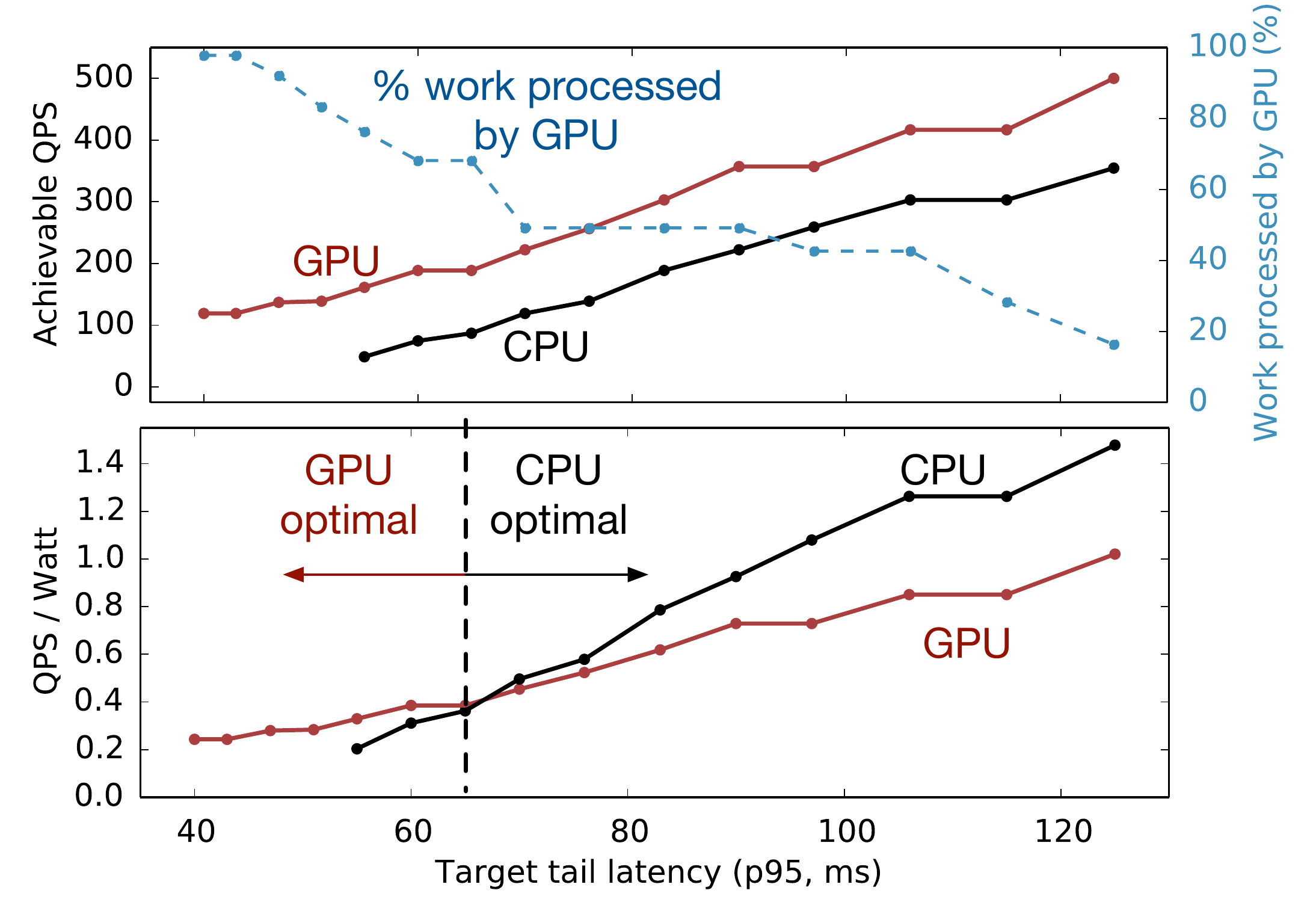}%
    \caption{ (Top) System throughput increases by scheduling queries across both CPUs and  GPUs.
    The percent of work processed by the GPU decreases at higher tail latency targets. 
    (Bottom) While QPS strictly improves, the optimal configuration based on QPS/Watt, varies based on the tail latency targets.
    GPUs are optimal at low tail latencies while CPUs provide better power efficiency at higher tail latency targets. }
    \label{fig:gpu_scheduling}%
    \vspace{-1em}
\end{figure}

\subsection{Leverage Parallelism with Specialized Hardware }
In addition to trading off request vs. batch-level parallelism, \Sched-GPU leverages additional parallelism by offloading recommendation inference queries to GPUs.

\textbf{Performance improvements.}
GPUs are often treated as throughput-oriented accelerators as compared to CPUs. However, in the context of personalized recommendation, 
we find that GPUs can unlock lower tail latency targets that could not be achieved by the CPU.
Figure~\ref{fig:gpu_scheduling}(a) illustrates the performance impact of scheduling requests across both CPUs and GPUs.
While the lowest achievable tail-latency targets for DLRM-RCM1 on CPUs is 57$ms$, GPUs can achieve a tail-latency target of as low as 41$ms$ (1.4$\times$ reduction).
This is a result of recommendation models exhibiting high compute and memory intensity, as well as the heavy tail of query sizes in production use cases (Figure~\ref{fig:query_size}).

Next, in addition to achieving lower tail latencies, parallelization across both the CPU and the specialized hardware continue to increase system throughput.
For instance, Figure~\ref{fig:gpu_scheduling}(a) shows that across all tail-latency targets, \Sched-GPU achieves higher QPS than \Sched-CPU.
This is as a result of the execution of the larger queries on GPUs, enabling higher system throughput.
Interestingly, the percent of work processed by the GPU decreases with higher tail latency targets.
This is due that, at a low latency target, \Sched-GPU optimizes system throughput by setting a low query size threshold and offloads a large fraction of queries to the GPU.
Under a more relaxed tail-latency constraint, more inference queries can be processed by the CMPs. 
This leads to a higher query size threshold for \Sched-GPU. 
At a tail latency target of 120$ms$, the optimal query size threshold is 324 and the percent of work processed by the GPU falls to 18\%.
As shown in Figure~\ref{fig:design_opt}(top), optimizing the query size threshold yields \Sched-GPU's system throughput improvements over the static baseline and \Sched-CPU across the different tail latency targets and recommendation models.



\textbf{Infrastructure efficiency implications.}
While GPUs can enable lower latency and higher QPS, power efficiency is not always optimized with GPUs as the specialized AI accelerator. For instance, Figure~\ref{fig:gpu_scheduling}(b) shows the QPS/Watt of both \Sched-CPU and \Sched-GPU for DLRM-RMC1, across a variety of tail latency targets.
At low tail latency targets, QPS/Watt is maximized by \Sched-GPU --- parallelizing queries across both CPUs and GPUs.
However, under more relaxed tail-latency targets, we find QPS/Watt is optimized by processing queries on CPUs only.
Despite the additional power overhead of the GPU, \Sched-GPU does not provide commensurate system throughput benefits over \Sched-CPU at higher tail latencies.

More generally, power efficiency is co-optimized by considering both the tail latency target and the recommendation model.
For instance, Figure~\ref{fig:design_opt}(b) illustrates the power efficiency for the collection of recommendation models across different tail latency targets.
We find that \Sched-GPU achieves higher QPS/Watt across all latency targets for compute-intensive models (i.e., NCF, WnD, MT-WnD) --- the performance improvement of specialized hardware outweighs the increase in power footprint.
Similarly, for DLRM-RMC2 and DIEN, \Sched-GPU provides marginal power efficiency improvement compared to \Sched-CPU.
On the other hand, the optimal configuration for maximizing power efficiency of DLRM-RMC1 and DLRM-RMC3 varies based on the tail latency target.
As a result, as shown in Figure~\ref{fig:design_opt}(b), in order to maximize infrastructure efficiency, it is important to consider a variety of recommendation uses cases, including model architecture and tail latency targets.

\section{Related Work}
While the system and computer architecture community has devoted significant efforts to characterize and optimize deep neural network (DNN) inference efficiency, relatively little work has explored running recommendation at-scale.

\textbf{DNN accelerator designs.} 
Currently-available benchmarks for DNNs primarily focus on FC, CNNs, and RNNs~\cite{fathom,zhu2018benchmarking,coleman2017dawnbench,deepbench,wei2019benchmarking}.
Building upon the performance bottlenecks, a variety of software and hardware solutions have been proposed to optimize traditional DNNs ~\cite{scnn, masr2019, eyeriss, eie, minerva, gonzalez2018pact,fletcher2019micro,fletcher2018micro,dadiannao,cambricon, sharma2016dnnweaver, rhu2016vdnn,diannao,du2015shidiannao,liu2015pudiannao,chi2016prime,shafiee2016isaac,kim2016neurocube,likamwa2016redeye,mahajan2016tabla,gao2017tetris,maxnvm,kwon2018beyond,choi2019prema,albericio2017bit}.
While the benchmarks and accelerator designs consider a variety of DNN use cases and systems, prior solutions do not apply to the wide collection of state-of-the-art recommendation models presented in this paper.
For example, recent characterization of Facebook's DLRM implementation demonstrates that DNNs for recommendation have unique compute and memory characteristics\cite{naumov2019dlrm, gupta2019architectural}.
These implementations are included, as DLRM-RMC 1-3, within \Infra. 
In addition, MLPerf, an industry-academic benchmark suite for machine learning, provides NCF as a training benchmark~\cite{mlperf}. 
NCF, however, is not continued in the latest release; MLPerf is developing a recommendation benchmark that is more representative of industry e-commerce tasks for the next submission round~\cite{mlperf-training,mlperf-inference}. 
In addition, a unique and very important aspect of the end-to-end infrastructure presented in the paper is taking into account the at-scale inference request characteristics (arrival rate and size), particularly important for recommendation.

\textbf{Optimizations for personalized recommendation.}
There are a few recent works that explored the design optimization opportunities for recommendation models.
For instance, TensorDimm proposes and evaluates a near memory processing solution for recommendation models similar to  DLRM-RMC 1-3 and NCF~\cite{kwon2019tensordimm}.
Ginart et al. and Shi et al.~\cite{ginart2019mixed, shi2019compositional} propose optimization techniques to compress embedding tables in recommendation models while maintaining the model accuracy.
In contrast, this paper optimizes the at-scale inference performance of a wider collection of recommendation models by considering the effect of inference query characteristics as well as tail latency targets specific to distinct use cases. 

\textbf{Machine learning at-scale.}
Finally, prior work has examined the performance characteristics and optimization techniques for ML running on at-scale, warehouse scale machines.
Sirius and DjiNN-and-Tonic explore the implications of ML in warehouse-scale computers~\cite{hauswald2015sirius, djinn}.
However, the unique properties of recommendation inference and query patterns have not been the focus of the prior work.
Li et al. ~\cite{li2016work} exploit task and data-level parallelism to meet SLA targets of latency critical applications i.e., Microsoft's Bing search and finance workloads.
Furthermore, recent work has open-sourced benchmarks for studying the performance implication of at-scale execution of latency critical datacenter workloads and cloud micro-services~\cite{delimitrou2019asplos,kasture2016tailbench}. 
In contrast, this paper provides an end-to-end infrastructure (\Infra) and design solutions (\Sched) specialized for at-scale recommendation inference.
\Infra provides an even baseline for state-of-the-art recommendation models. 
It models real-time query patterns, representative of the distinct working set size distribution in production datacenter fleets. 
The unique characteristics lead to the design of \Sched, providing significant performance improvement for at-scale recommendation --- an important yet understudied class of AI inference. 

%


\section{Conclusion}
Given the growing ubiquity of web-based services that use recommendation algorithms, such as search, social-media, e-commerce, and video streaming, deep learning-based personalized recommendation comprises the majority of AI inference capacity and cycles in production datacenter.
We propose \Infra, an extensible infrastructure to study a variety of at-scale recommendation inference.
The infrastructure comprises eight state-of-the-art recommendation models, SLA targets, and query patterns.
Built upon this framework, \Sched~exploits the unique characteristics of at-scale recommendation inference in order to optimize system throughput under a strict tail latency constraint.
Across eight recommendation models and under a variety of SLA targets, we demonstrate that \Sched~improves system throughput by 2$\times$.
In addition to evaluating \Sched~on \Infra, the design optimizations are evaluated in a real production datacenter demonstrating similar performance benefits.
Finally, through judicious optimizations, \Sched~can leverage additional parallelism by offloading queries across CPUs and specialized AI hardware in order to achieve higher system throughput and infrastructure efficiency.



\section{Acknowledgements}
We would like to thank Cong Chen and Ashish Shenoy for the valuable feedback and numerous discussions on the at-scale execution of personalized recommendation systems in Facebook's datacenter fleets. The collaboration leads to insights which we use to refine the proposed design presented in this paper. It also results in design implementation, testing, and evaluation of the proposed idea for production use cases.


\bibliographystyle{ieeetr}
\bibliography{main}

\end{document}